\documentstyle[12pt]{article}
\def\beq{\begin{equation}}
\def\eeq{\end{equation}}
\def\beqa{\begin{eqnarray}}
\def\eeqa{\end{eqnarray}}

\begin{document}

\begin{center}
{\bf FINAL STATE INTERACTIONS IN THE DECAY OF HEAVY QUARKS\footnote{
Talk presented at the 3rd German-Russian Workshop on Heavy Quark
Physics, Dubna, May 1996} \\}

\vspace*{1cm}
JOHN F. DONOGHUE\\
{\it Department of Physics and Astronomy,\\} 
{\it University of Massachusetts, Amherst MA 01003 USA}
\end{center}
\vspace*{1.5cm}
\begin{abstract} 
{\small I discuss some
recent results on the systematic behavior of final state rescattering
which makes use of the limit of a heavy B meson. The results suggest
that soft final state interactions do not disappear in the large $m_B$
limit. Soft and hard final state phases can both contribute to CP
violating asymmetries in B decay. The way that the soft phases occur
is interesting theoretically and suggests the violation of local
quark-hadron duality.}
\end{abstract}
\vspace*{1.5cm}
Some of the CP violating asymmetries that can occur in the Standard
Model require the presence of strong final state interaction phases.
Very little in known about final state interactions at the high energies
that are relevant for B decay. What I will describe in this talk does
not ``solve'' the problem of these interactions. However, my
collaborators (Eugene Golowich, Alexey Petrov and Jo\~ao Soares) and I
have obtained some insights that at least taught me something about
this frustrating topic$^1$. I would like to share these with you.
This has taken us on a path into
unfashionable but surprisingly interesting physics and may shed new light
on weak decays. As is now common, we use the mass of the B meson as an
organizing parameter and consider the limit where the mass is very
large. We can show that soft final state interaction survive in this
limit and can isolate their leading cause, although 
in the end we cannot provide a
specific number describing the magnitude of the phase.

Our main conclusions rest on a few simple facts. These are: 

1) The Unitarity Relation. 
Final state interactions in B decay involve the rescattering of
physical final state particles.  Unitarity of the
${\cal S}$-matrix, ${\cal S}^\dagger {\cal S} = 1$, implies 
that the ${\cal T}$-matrix, ${\cal S} = 1 + i {\cal T}$, obeys 
\beq
{\cal D}isc~{\cal T}_{B \rightarrow f} \equiv {1 \over 2i} 
\left[ \langle f | {\cal T} | B \rangle - 
\langle f | {\cal T}^\dagger | B \rangle \right] 
= {1 \over 2} \sum_{I} \langle f | {\cal T}^\dagger | I \rangle 
\langle I | {\cal T} | B \rangle \ \ .
\label{unit}
\eeq
Of interest are all physical intermediate states which can scatter
into the final state $f$.  

2)The optical theorem.
  The optical theorem
relates the forward invariant amplitude ${\cal M}$ to the total 
cross section, 
\beq
{\cal I}m~{\cal M}_{f\to f} (s, ~t = 0) = 2 k 
\sqrt{s} \sigma_{f \to {\rm all}} \sim s \sigma_{f \to {\rm all}} \ \ ,
\label{opt}
\eeq
where $s$ is the squared center-of-mass energy and $t$ is the squared
momentum transfer. 

3) The exponential fall-off in momentum transfer.
Soft hadronic interactions are categorized by a limited momentum
transfer, and all of the high energy hadronic reactions have an
exponential damping of the form 

\beq
{\cal M}(s,t) \simeq f(s) e^{bt} \ \ .
\label{fall}
\eeq
Recall that t is negative. This damping limits the momentum transfer to be of
order 0.5 $GeV$.

4) The measured cross sections.
The asymptotic total cross sections are known
experimentally to rise slowly with energy.  All known cross sections
can be parameterized by fits of the form$^2$ 
\beq
\sigma (s) = X \left({s\over s_0}\right)^{0.08} 
+ Y \left({s\over s_0}\right)^{-0.56} \ \ ,
\label{pl}
\eeq
where $s_0 = {\cal O}(1)$~GeV is a typical hadronic scale.  
Combined with the optical theorem, this implies that
 the imaginary part of the forward elastic scattering amplitude 
rises asymptotically as $s^{1.08}$.  

This growth with s is an important ingredient in our results. (Note
that the emphasis is on the factor of $s^1$; the extra factor of
$s^{0.08}$ that occurs repeatedly below is not particularly
important.) It is a
surprising feature in that it cannot be generated by a perturbative
mechanism at any finite order.  In particular, standard 
calculations based on
the quark model or perturbative $QCD$ would completely miss this
feature.

These indisputable facts can be combined to show that final state
rescattering does not disappear in the limit of large $m_B$.
In order to arrive most simply at this result, 
let us consider first only the imaginary part of the amplitude.
Building in 
the features described above one has
\beq
i{\cal I}m~{\cal M}_{f\to f} (s,t) \simeq i \beta_0 \left( {s \over s_0}
\right)^{1.08} e^{bt} \ \ .
\label{ampim}
\eeq
It is then an easy task to calculate the contribution 
of the imaginary part of the elastic amplitude to the
unitarity relation for a final state $f = a + b$ with kinematics 
$p_a' +  p_b' = p_a +  p_b$ and $s = (p_a + p_b )^2$. We find 
\beqa
{\cal D}isc~{\cal M}_{B \to f} &=&
{1 \over 2} \int {d^3p_a' \over (2\pi)^3 2E_a'}
{d^3p_b' \over (2\pi)^3 2E_b'}
(2 \pi)^4 \delta^{(4)} (p_B - p_a' - p_b') \\ &\cdot & -i\beta_0 
\left( {s \over s_0} \right)^{1.08} e^{b(p_a - p_a')^2} 
{\cal M}_{B \rightarrow f} \nonumber \\
&=&-{i \over 32\pi}\int d(cos\theta ) e^{-{bs \over 2}(1-cos\theta)}
\beta_0\left( {s \over s_0} \right)^{1.08} {\cal M}_{B \rightarrow f}
\nonumber \\ 
&=& - {1\over 16\pi} {i\beta_0 \over s_0 b}\left( {m_B^2 \over s_0} 
\right)^{0.08} {\cal M}_{B \rightarrow f} \ \ ,
\label{mess}
\eeqa
where $t = (p_a - p_a')^2 \simeq 
-s(1 - \cos\theta)/2$ and we have taken $s = m_B^2$. 
There are two competing effects that are important in this result. The
first is a kinematic suppression of soft final state interactions
because of the limited angular region corresponding to the soft
region. The integration over the angle involving the direction of the 
intermediate state is seen to introduce a suppression factor 
to the final state interaction of $s^{-1} = m_B^{-2}$.  
This is because the soft final state rescattering can take place only 
if the intermediate state has a transverse momentum $p_\perp 
\le 1$~GeV with respect to the final particle direction.  This would 
naively suggest a result consistent with conventional 
expectations, {\it i.e.} an FSI which falls as $m_B^{-2}$.  However,
the second feature is the fact, mentioned above, 
that the forward scattering amplitude {\it grows} with a
power of $s$ which overcomes this suppression and leads to elastic 
rescattering which does not disappear at large $m_B$.  

In fact, we can make a more detailed estimate of elastic 
rescattering because the phenomenology of high energy 
scattering is well accounted for by Regge theory$^3$.  
Scattering amplitudes are described by the exchanges of 
Regge trajectories (families of particles of differing spin) 
which lead to elastic amplitudes of the form 
\beq
{\cal M}_{f \rightarrow f} = \xi \beta (t) 
\left( {s \over s_0} \right)^{\alpha (t)} e^{i \pi \alpha(t)/2}
\label{regge}
\eeq
with $\xi = 1$ for charge conjugation $C=+1$ and 
$\xi = i$ for $C=-1$. Each such trajectory is described by a straight line, 
\beq
\alpha (t) = \alpha_0 + \alpha' t \ \ .
\label{traj}
\eeq
The leading trajectory for high energy scattering is the 
Pomeron, having  $C=+1, \alpha_0 \simeq 1.08$ and 
$\alpha' \simeq 0.25$~GeV$^{-2}$. 
Using known features of Pomeron physics and taking 
$s = m_B^2 \simeq 25~{\rm GeV}^2$, we obtain for the Pomeron
contribution 
\beq
{\cal D}isc~{\cal M}_{B \to \pi\pi}|_{\rm Pomeron} = -i\epsilon
{\cal M}_{B \to \pi\pi} \ \ ,
\label{despite}
\eeq
where we find from our computation, 
\beq
\epsilon \simeq 0.21 \ \ .
\label{eps}
\eeq
The simplest conclusion from this calculation is that final state
interactions survive in the large $m_B$ limit and are reasonably
large. 

This calculation also tells us more: it requires that the inelastic
channels be at least equally important, and that they are the key 
to the origin
of the final state phases. This is due to the fact that the elastic
effect calculated  above is purely imaginary. In the limit of
T-invariance, the discontinuity ${\cal D}isc {\cal M}$ is a real
number up to an irrelevant rephasing invariance of the B-state. The
factor of $i$ in the elastic amplitude must be removed by the effects
of the inelastic rescattering channels. This implies that inelastic
rescattering cannot be vanishingly small and must share the same power
behavior in $m_B$ as the elastic amplitude. At a physical level this
is not at all surprising since a two body initial state scatters
primarily inelastically at high energy.
In fact, the elastic calculation implies even more, in that the
inelastic channels can be considered systematically {\it larger} than
the elastic channel. We have a 
 ${\cal T}$-matrix element ${\cal T}_{ab
\to ab} = 2 i \epsilon$, which directly gives ${\cal S}_{ab \to ab} = 1- 2 
\epsilon$. However, the constraint of the ${\cal S}$-matrix be unitary 
can be shown to imply that the 
off-diagonal elements must be ${\cal O}(\sqrt{\epsilon})$. 
Since $\epsilon$ is 
approximately ${\cal O}(m_B^0)$ in powers of $m_B$ and numerically 
$\epsilon < 1$, the inelastic amplitude must also be ${\cal
O}(m_B^0)$ and of magnitude $\sqrt{\epsilon} > \epsilon$.  
  Therefore, the
presence of inelastic effects is seen to be necessary.  

It is possible to illustrate 
the systematics of inelastic scattering by means of a simple 
two-channel model.  This pedagogic example involves a two-body final state 
$f_1$ undergoing elastic scattering and a final state $f_2$ which is meant to
represent `everything else'.  We assume that the elastic amplitude 
is purely imaginary.  Thus, the scattering can be described 
in the one-parameter form 
\beq
 S=  \pmatrix{ 1-2\epsilon &  
 2i{\sqrt{\epsilon}} \cr
             2 i {\sqrt{\epsilon}} & 1-2\epsilon \cr} \ ,\qquad \qquad 
 T = \pmatrix{2 i\epsilon &  2{\sqrt{\epsilon}} \cr
             2{\sqrt{\epsilon}} & 2 i \epsilon \cr} \ \ ,
\label{matr1}
\eeq
(These are approximate forms valid to order $\epsilon$. It is not hard
to use exactly unitary forms, but I find it instructive to
explicitly display the powers of $\epsilon$.)
 The unitarity relations become
\begin{eqnarray}
{\cal D}isc ~{\cal{M}}_{B \to f_1} = - i \epsilon {\cal{M}}_{B \to f_1} +
{\sqrt{\epsilon}} {\cal{M}}_{B \to f_2} \ \ ,\nonumber \\
{\cal D}isc~ {\cal {M}}_{B \to f_2} = {\sqrt{\epsilon}} 
{\cal{M}}_{B \to f_1} - i \epsilon {\cal{M}}_{B \to f_2} \ \ 
\label{big}
\end{eqnarray}
If, in the limit $\epsilon \to 0$, the decay amplitudes become the real numbers 
${\cal{M}}_1^0$ and ${\cal{M}}_2^0$, these equations are solved by
\beq
{\cal{M}}_{B \to f_1} =  {\cal{M}}_1^0 + i {\sqrt{\epsilon}}
{\cal{M}}_2^0  \ , \qquad 
{\cal{M}}_{B \to f_2} =  {\cal{M}}_2^0 + i {\sqrt{\epsilon}}
{\cal{M}}_1^0 \ \ .
\label{soln}
\eeq
As a check, we can insert these solutions back into Eq.~(\ref{big}). 
Upon doing so and bracketing contributions from ${\cal{M}}_{B \to
f_1}$ and ${\cal{M}}_{B \to f_2}$ separately, we find 
\begin{equation}
{\cal D}isc~{\cal{M}}_{B \to f_1} = {1\over 2}\left[ \bigg( -2i\epsilon 
{\cal M}^0_{B\to f_1} 
+ {\cal O}(\epsilon^{3/2}) \bigg) + \bigg( 2\sqrt{\epsilon} 
{\cal M}^0_{B\to f_2} + 2i\epsilon {\cal M}^0_{B\to f_1} \bigg) \right]\ \ .
\label{check}
\end{equation}
The first of the four terms comes 
from the elastic channel $f_1$ and is seen to be 
canceled by the final term, which arises from the inelastic channel
$f_2$.  The third term is dominant, being  ${\cal O}(\sqrt{\epsilon})$, 
and comes from the inelastic channel.  

In this example, we have seen that the phase is given by the inelastic
scattering with a result of order 
\begin{equation}
\frac{ {\cal I}m~ {\cal{M}}_{B \to f}}{{\cal R}e~ {\cal{M}}_{B \to f}} \sim 
\sqrt{ \epsilon}~ \frac{{\cal{M}}_2^0}{{\cal{M}}_1^0} \ \ .
\end{equation}
Clearly, for physical $B$ decay, we no longer 
have a simple one-parameter ${\cal S}$ matrix.  However, the main
feature of the above result is expected to remain --- that 
inelastic channels {\em cannot} vanish because they 
are required to make the discontinuity real and that 
the phase is systematically of order $\sqrt{\epsilon}$ from these
channels.  Of course, with many channels, cancellations or enhancements are 
possible for the sum of many contributions. However the generic expectation
remains --- that inelastic soft final-state-rescattering arising
from Pomeron exchange will generate a phase which does not vanish in
the large $m_B$ limit.

What about nonleading effects? It is not hard to see that 
these may be significant at the physical values of $m_B$. For example, 
the fit to the $\bar p p$ total cross section is
\begin{equation}
\sigma ( p \bar p ) = \Bigl [ 22.7 \Bigl ( \frac{s}{s_0} \Bigr )^{0.08} +
140 \Bigl ( \frac{s}{s_0} \Bigr )^{-0.56} \Bigr ] ~~ (mb)
\end{equation}
with $s_0 = 1~{\rm GeV}^2$. At $s=(5.2~{\rm GeV})^2$, the nonleading 
coefficient is a factor of six larger that the leading effect,
effectively compensating for the $s^{-0.56}=m_B^{-1.12}$ suppression.
The subleading terms are then comparable in  
the elastic forward $\bar p p $ scattering amplitude.  There 
are several next-to-leading trajectories, both those with $C=-1$ 
($\rho(770)$ $\&$ $\omega(782)$ trajectories) and those with $C=+1$ 
($a_2 (1320)$ $\&$ $f_2(1270)$ trajectories).  Roughly, these have 
$\alpha_0 \simeq 0.44$, $\alpha' \simeq 0.94$~GeV$^{-2}$ and lead 
collectively to the $s^{-0.56}$ 
dependence in the asymptotic cross section of Eq.~(\ref{pl}). 
If we estimate the $\beta$ coefficient of
the $\rho$ trajectory in $\pi \pi$ by relating it to $\bar p p$ via a factor 
of $\beta_{\pi \pi} \simeq 4 \beta_{\bar p p} $ and then perform the 
integration over the intermediate state momentum we find
\begin{equation}
Disc~{\cal{M}}_{B \to \pi\pi} \bigg|_{\rho - {\rm traj}} = 
i \epsilon_\rho {\cal{M}}_{B \to \pi\pi} \ \ ,
\end{equation}
with $\epsilon_\rho \simeq 0.11 - 0.05~i$.  It is likely that the 
$f_2 (1270)$ trajectory could be somewhat larger, as it is in 
$\bar p p$ and $\pi p$ scattering.  

In addition to the soft physics described above, one may expect that
hard physics also may generate final state interaction phases. Hard
physics is characterized by larger momentum transfer and is best 
described by the exchanges of quarks and gluons. The final state
interactions then correspond to rescattering of intermediate states
which are modeled by
on-shell quarks and gluons. These arise as imaginary parts in the Feynman
diagrams relevant for the decays.  It might be thought that such
phases are always of order $\alpha_s$, but this need not be always
true. For example, the best known counterexample occurs in the penguin
diagram. Here the physical intermediate state is the 
$q\bar{q}$ (with $q=u,c$) in the loop of the penguin
diagram which can go onshell, yielding an imaginary part to the diagram. The
hard rescattering is the transition of this intermediate state into
the final $q\bar{q}$ pair (eq. $c\bar{c} \to s\bar{s}$) through a
gluon. Both the real part of the penguin diagram and the imaginary
part are of order $\alpha_s$ and therefore the phase occurs at the
zeroth order in $\alpha_s$. A similar situation can occur even in the
W-exchange class of operators, for the operator that is often called
the color-octet operator. For this case, the quarks that are to emerge
in a particular final hadron (eg. a $c\bar{d}$ for a $D^+$)
occur in the operator in a color octet
combination. Therefore in a factorization scheme, the matrix element
would vanish since the quarks in the hadron are in a color singlet. 
However, with the exchange of a gluon the matrix element
can be non-zero. The same gluon intermediate state can generate an
imaginary part to the amplitude, so that again both the real and
imaginary parts of the diagram can end up being of the same order in
$\alpha_s$. (This case is not as clear as that of the penguin diagram
because it assumes the the real part of these operator matrix elements
is generated in a perturbative fashion, which has not been carefully
explored.) The reverse situation might occur for what is called the
color singlet operator, where the quarks are in a singlet state
allowed by the factorization hypothesis. Here the intermediate states
with an imaginary part from single gluon exchange in the final state
are forbidden by the color structure, and the lowest gluonic
rescattering occurs with two gluon exchange, leading to a hard phase
naively of order $\alpha_s^2$.

Both the soft and hard phases can contribute to some of the CP
violating asymmetries. As an example consider the decays $B \to K
\pi$. Both W-exchange diagrams and penguin diagrams can contribute to
the amplitude, and these diagrams have different weak phases$^{4,5}$. 
What is
then required for an observable CP violating asymmetry is for these
two sets of diagrams to also have different final state interaction
phases. The asymmetry is generated by an interference of the two
types of phases, with 
\begin{eqnarray}
 {\cal{M}}(B\to K \pi) & = & A_w e^{i\phi_w}e^{i\delta_w} + A_p e^{i\phi_p}
e^{i\delta_p} \ \ \nonumber \\
 {\cal{M}}(\bar{B}\to \bar{K} \pi) & = & A_w e^{-i\phi_w}e^{i\delta_w} 
+ A_p e^{-i\phi_p}
e^{i\delta_p} 
\end{eqnarray}
\noindent leading to 
\beq
\Delta \Gamma \sim A_w A_p sin(\phi_w - \phi_p) sin(\delta_w -
\delta_p)  \ \ .
\eeq
\noindent The hard interactions in the penguin diagram can generate
the required FSI phase, as described above. What is perhaps not as
obvious is that soft interactions can also generate this phase
difference. This occurs because the W-exchange and penguin diagrams
will in general populate the elastic and inelastic channels in
different ratios. Even if the soft rescattering is the same, this
would lead to different FSI phases for the two classes of diagrams
(see Eq. (16)). Both the hard and soft contributions are of the same
order in the parameters ($m_B$ and $\alpha_s$) which we are using 
to characterize the transitions. This means that one cannot simply
calculate the final state phase difference by a perturbative
calculation of the penguin diagram. 

The above situation is also of interest
theoretically, as it is an example of a violation of the loose notion
of local quark hadron duality, which would have implied that a
calculation of the process at the quark-gluon level would have given
the hadron level answer even for exclusive quantities when suitably
averaged. The soft final state phases are quantities that would not
arise in conventional quark level calculations, and hence are always
outside of the realm of local quark hadron duality. This has occurred
because of the growth of the forward amplitude with s. The soft
interactions will limit
the accuracy of models which are based on quark level ideas ignoring
final state interactions. It is also in intriguing possibility that
the assumption of local quark hadron duality can be questioned in
other aspects of weak decays also. This is a topic which deserves more
study. The final state interactions in general are a subject about
which little is understood. The scaling properties described by our
study give at least a little insight into the physics of these
interactions.

\vspace{0.5cm}

{\bf { References}}\\

\vspace{0.25cm}

\noindent 1) J.F. Donoghue, E. Golowich, A.A. Petrov and J.M. Soares,
hep-ph/9604283  \\

\noindent 4) For example, see Section~5 of Chapter~XIV in 
J.F. Donoghue, E. Golowich and B.R. Holstein, {\it Dynamics of 
the Standard Model}, (Cambridge University Press, Cambridge, England 1992).\\
 
\noindent 5) Lincoln Wolfenstein, {\it Phys. Rev.} {\bf D43} (1991) 151. \\
 
\noindent 2) P.V. Landshoff, in {\it QCD - 20 Years Later}, 
(World Scientific, Singapore 1993).\\

\noindent 3) For example, see P.D.B. Collins, {\it Introduction to 
Regge Theory and High Energy Physics}, (Cambridge University Press, 
Cambridge, England 1977).\\

\end{document}